\magnification=\magstep1
\centerline{\bf On the Minimum Energy Configuration of a Rotating 
Barotropic Fluid}
\bigskip
\bigskip
\centerline{Ramesh Narayan and James E. Pringle}
\bigskip
\bigskip
\noindent
{\bf 1. Introduction}
\medskip

In a recent paper (astro-ph/0207561, posted on July 25, 2002), Fromang
\& Balbus (2002, hereafter FB02) claim that the minimum energy
configuration of a rotating fluid depends critically on whether or not
any part of the fluid rotates supersonically.  They state that ``a
uniformly rotating barotropic fluid in an external potential attains a
true energy minimum if and only if the rotation profile is everywhere
subsonic.  If regions of supersonic rotation are present, fluid
variations exist that could take the system to states of lower
energy.''

We present here a simple counter-example which demonstrates that the
stablity criterion presented in FB02 is incorrect.

\bigskip\noindent
{\bf 2. Minimum Energy Equilibria}
\medskip

We first repeat part of their analysis.  Following FB02, we consider a
barotropic fluid in which the pressure is a unique function of the
density.  To further simplify matters, and to be consistent with the
analysis in the next section, we consider a polytropic equation of
state in which the pressure $P$ and the density $\rho$ are related as
$$
P=K\rho^\gamma. \eqno (1) 
$$
Here $K$ is a constant throughout the fluid and is not allowed to vary
in any perturbation. Thus we exclude thermal effects. Furthermore, in
line with FB02, we consider only perturbations that are adiabatic,
which excludes convective instabilties.

For such a fluid, the internal energy per unit volume $\epsilon$, the
enthalpy per unit mass $H$, and the adiabatic sound speed $a$, are
given by
$$ 
\epsilon = {P\over
\gamma-1}, \qquad H = {\gamma \over \gamma-1} {P\over \rho}, \qquad
a^2 = {\gamma P \over \rho}. \eqno (2)
$$

We assume that the mass $M$ and the total angular momentum $J$ of the
system are held fixed:
$$
M = \int \rho dV = {\rm constant}, \eqno (3)
$$
$$ 
J = \int \rho Rv dV = {\rm constant}, \eqno (4) 
$$ 
where $R$ is the cylindrical radius as measured perpendicular to the
rotation axis and $v(R)$ is the azimuthal velocity of the gas at
radius $R$.  Note that, as do FB02, we take account throughout only of
the azimuthal component of the velocity. The implication of this is
that we restrict stability considerations to axi-symmetric
perturbations alone.

We are interested in minimizing the energy, $E$,  of the system,
$$
E = \int \left ({1\over2}\rho v^2 + \epsilon + \rho \phi
\right ) dV, \eqno (5)
$$
where $\phi$ is the (time-independent) external potential in which the
fluid is confined.  (Following FB02, we neglect self-gravity of the
fluid.)  The minimiziation of $E$ subject to the constraints (3) and
(4) is a standard problem that can be solved by variational methods.
We minimize
$$
I = E- \zeta M - \Omega J = \int \left({1\over2}\rho v^2 + \epsilon +
\rho \phi - \zeta \rho - \Omega \rho R v \right ) dV, \eqno (6)
$$
where $\zeta$ and $\Omega$ are Lagrange multipliers.  To find an
extremum of the energy, we set the first order variation of $I$
with respect to changes in $\rho (R)$ and $v(R)$ to zero:
$$
\delta I = \int \left[ \left( {1\over2}v^2 -\Omega Rv
+H+\phi-\zeta\right)\delta\rho + \rho(v-\Omega R)\delta v
\right] dV = 0, \eqno (7)
$$
where we have used equation (1) to set $d\epsilon/d\rho=H$.  Since
$\delta I$ should vanish for arbitrary choices of $\delta \rho(R)$ and
$\delta v(R)$, the coefficients of $\delta \rho(R)$ and $\delta v(R)$
in the integrand of (7) should individually vanish at each $R$.
This gives the following two conditions:
$$
v=\Omega R, \eqno (8)
$$
$$
-{1\over2}\Omega^2R^2 + H + \phi = \zeta = {\rm constant}. \eqno (9)
$$
The first condition states that a polytropic fluid in an
energy-extremum configuration must be in a state of uniform rotation,
a well-known result (e.g. Landau \& Lifshitz).  The second condition
is nothing other than the condition of hydrostatic balance, as can be
verified by taking the gradient of equation (9).  We thus find that a
polytropic fluid with fixed $M$ and $J$ and with an extremum in $E$
has uniform rotation and is in hydrostatic equilibrium.

The analysis is identical to that in FB02, with the sole difference
that we have specialized to the case of a polytropic equation of
state.  Following the above steps, FB02 proceed to take second
differences of the energy, on the basis of which they reach the
conclusions stated above in \S1.  We do not repeat their analysis
since their final conclusions are incorrect, as we now demonstrate by
means of a counter-example.

\bigskip\noindent
{\bf 3. A Simple Counter-Example}
\medskip

We consider a polytropic fluid rotating around the $z$ axis in a
quadratic potential
$$
\phi (R,z) = {1\over2}\Omega_K^2(R^2+z^2). \eqno (10)
$$
We assume that the fluid is in an energy extremum state which, by the
analysis of the previous section, means that the fluid rotates with a
constant angular velocity $\Omega$ and is in hydrostatic equilibrium,
satisfying equation (9).  Let us assume that the fluid extends from
$R=0$ to a maximum radius $R_0$ on the equatorial plane.  (Such a
maximum radius always exists for $\gamma>1$.)  At the point
$(R,z)=(R_0,0)$, the enthalpy of the gas vanishes.  This enables us to
determine the constant $\zeta$ in equation (9), using which we can
show that
$$
H(R,z) = {1\over2}(\Omega_K^2-\Omega^2)\left[ R_0^2 - R^2 -
{\Omega_K^2\over(\Omega_K^2-\Omega^2)}z^2\right]. \eqno (11)
$$
Not surprisingly, the fluid takes the form of a flattened spheroid,
with equatorial radius $R_0$ and ``vertical'' radius
$$
Z_0 = \left({\Omega_K^2-\Omega^2\over\Omega_K^2}\right)^{1/2}
R_0. \eqno (12)
$$

From the expression for $H$, we may calculate the density $\rho$ of
the fluid,
$$
\rho(R,z) = \left[{(\gamma-1)H\over\gamma K}\right]^n, \eqno (13)
$$
where we have defined the polytropic index $n$ in the usual way:
$$
n = {1\over \gamma-1}. \eqno (14)
$$
The vertically integrated surface density of the fluid is then
$$
\Sigma(R) = C (R_0^2-R^2)^{n+1/2}, \eqno (15)
$$
$$
C = {\Gamma({1/2})\Gamma(n+1)\over\Gamma(n+3/2)}
\left({\gamma-1\over2\gamma K}\right)^n
{(\Omega_K^2-\Omega^2)^{n+1/2}\over\Omega_K}. \eqno (16)
$$

Through equation~(16), the angular velocity $\Omega$ defines the
parameter $C$. Thus this simple model has two free parameters, $R_0$
and $\Omega$.  For a given value of $\Omega$ we may determine the
radius $R_0$ by applying the mass constraint
$$
M = \int \Sigma(R)2\pi RdR = 
{\pi\over (n+3/2)} C R_0^{2n+3}. \eqno (17)
$$
We then determine $\Omega$ by applying the angular momentum
constraint,
$$
J = \int \Sigma(R) \Omega R^2 2\pi RdR = 
{1\over(n+5/2)}M\Omega R_0^2. \eqno (18)
$$

From equations~(16) and~(17) we see that for a given mass, $R_0^2$
increases with increasing $\Omega$, and thus from equation~(18) it is
clear that $J$ is also a monotonically increasing function of
$\Omega$.  We conclude that for given values of $M$ and $J$, there is
one and only one energy extremum configuration.

All possible fluid configurations in the model have a positive energy,
since all the terms in equation (5) are positive.  Therefore, we are
guaranteed that the system has a well-defined global energy minimum.
Since we have shown that for given values of mass $M$ and angular
monentum $J$, there is only one configuration that is an energy
extremum, that particular configuration must correspond to the
absolute energy minimum state and must be stable.

In this simple analytical model, the orbital velocity $v(R)$ increases
monotonically with radius $R$, from $v=0$ at $R=0$ to a maximum at
$R=R_0$, while the sound speed $a(R,z)$ decreases from a maximum at
$R=0, z=0$ to $a=0$ all over the surface (and specifically at the
outer edge $R = R_0, z=0$):
$$
v(R) = \Omega R, \qquad a^2(R,z) = (\gamma-1)H(R,z)
\propto \left(1 - {R^2\over R_0^2} - {z^2 \over Z_0^2}\right). \eqno (19)
$$
It is evident that the model includes both subsonic and supersonic
zones, and should therefore be unstable according to the criterion
given by FB02.  And yet, as proved above, the fluid is in an energy
minimum state and is stable.  The model is thus an explicit
counter-example to the result claimed in FB02 --- supersonic rotation
does {\it not} imply that the fluid has neighboring states with a
lower energy.

\bigskip\noindent
{\bf 4. A Final Comment}

In astrophysics, one is generally interested in fluids with free
surfaces at zero pressure rather than fluids confined between walls.
Since FB02 claim that the only stable barotropic configurations are
those that rotate uniformly {\it and} are subsonic at all radii, it is
interesting to ask whether such fluid configurations are at all
possible in the first place.  The answer is that, if we leave out the
strictly non-rotating case ($\Omega=0$) and if we restrict ourselves
to fluids with polytropic indices $n>0$, there are no isolated
configurations that meet FB02's requirements.  (We thank Martin Rees
for pointing this out.)

For instance, if the fluid extends to an infinite radius (as can
happen for an isothermal gas), then the rotation velocity
$v(R)\to\infty$ at large $R$ whereas the sound speed remains finite.
The fluid at large radius is thus guaranteed to be supersonic.  On the
other hand, if the fluid has an edge at a finite radius, as in the
simple model described in \S3, then $v$ is finite at the edge whereas
the sound speed goes to zero.  Once again, the fluid near the outer
edge rotates supersonically.

Thus, all uniformly rotating configurations for a wide class of fluids
have supersonic motions.  Since, surely, all of these fluids must
possess stable minimum energy equilibria, the result claimed by FB02
must be incorrect.  The example discussed in \S3 is a specific example
that explicitly demonstrates this.

\bye